\documentclass[bimj,fleqn]{w-art}
\usepackage{times}
\usepackage{w-thm}
\usepackage[authoryear]{natbib}
\usepackage{mathtools}

\usepackage[utf8]{inputenc}
\usepackage{amsmath}
\usepackage{amsfonts}
\usepackage{amssymb}
\usepackage{multirow}
\usepackage{amsthm}
\usepackage{graphicx}
\usepackage{placeins}
\usepackage{courier}

\usepackage{rotating}
\usepackage{tabularx}
\usepackage{multirow}

\usepackage{array}
\newcolumntype{L}[1]{>{\raggedright\let\newline\\\arraybackslash\hspace{0pt}}m{#1}}
\newcolumntype{C}[1]{>{\centering\let\newline\\\arraybackslash\hspace{0pt}}m{#1}}
\newcolumntype{R}[1]{>{\raggedleft\let\newline\\\arraybackslash\hspace{0pt}}m{#1}}

\makeatletter
\newsavebox\myboxA
\newsavebox\myboxB
\newlength\mylenA

\newcommand*\xoverline[2][0.75]{%
	\sbox{\myboxA}{$\m@th#2$}%
	\setbox\myboxB\null
	\ht\myboxB=\ht\myboxA%
	\dp\myboxB=\dp\myboxA%
	\wd\myboxB=#1\wd\myboxA
	\sbox\myboxB{$\m@th\overline{\copy\myboxB}$}
	\setlength\mylenA{\the\wd\myboxA}
	\addtolength\mylenA{-\the\wd\myboxB}%
	\ifdim\wd\myboxB<\wd\myboxA%
	\rlap{\hskip 0.5\mylenA\usebox\myboxB}{\usebox\myboxA}%
	\else
	\hskip -0.5\mylenA\rlap{\usebox\myboxA}{\hskip 0.5\mylenA\usebox\myboxB}%
	\fi}
\makeatother

\newcommand{\overX}{\xoverline[0.7]{X}}

\begin{document}
\DOIsuffix{bimj.200100000}
\Volume{52}
\Issue{61}
\Year{2010}
\pagespan{1}{}
\keywords{SNP Array; Change-Point; Copy-number; SaRa;segmentation;\\
}  

\title[Running title]{Screening and merging algorithm for the detection of copy-number alterations.}
\author[Pinheiro {\it{et al.}}]{Murilo S. Pinheiro} 
\author[]{Benilton S. Carvalho }
\author[dd]{Aluísio S. Pinheiro\footnote{Corresponding author: {\sf{e-mail: pinheiro@ime.unicamp.br}}, Phone: (+55)(19) 3521-6080, Fax: (+55)(19) 3521-6080}}
\address{Depart of Statistics, IMECC University of Campinas, Sérgio Buarque de Holanda, 651 13083-855. Campinas SP Brazil.}
\Receiveddate{zzz} \Reviseddate{zzz} \Accepteddate{zzz} 
	
\begin{abstract}We call {\it change-point problem} (CPP) the identification of changes in the probabilistic behavior of a sequence of observations. Solving the CPP involves detecting the number and position of such changes. In genetics the study of how and what characteristics of a individual's genetic content might contribute to the occurrence and evolution of cancer has fundamental importance in the diagnosis and treatment of such diseases and can be formulated in the framework of chage-point analysis. In this article we propose a modification to a existing method of segmentation with the objective of producing a algorithm that is robust to a variety of sampling distributions and that is adequate for more recent method of accessing DNA copy-number which might require a restriction on the minimum length of a altered segment.
\keywords{beta mixture; EM algorithm; microarray data; change-point analysis; CUSUM}
\end{abstract}
	
\maketitle    

\section{Introduction to Change-Point Problems and their Role in Genetics}
The problem of identifying a change in the probabilistic behavior of a sequence of observations is usually called the {\it change-point problem} (CPP). Solving the CPP involves detecting the number and position of such changes. It is common to assume that the change-point structure is such that stochastic properties of the sequence are approximately piecewise constant.

Let $\{X_1,\ldots,X_n\}$ be a sequence of independent random variables (i.r.v.). We say that $\{X_1,\ldots,X_n\}$ is segmented with respect to the sequence of {\it change-points} $1<\tau_{1,n}<\ldots<\tau_{s,n}\leq n$ if, and only if, there are $s+1$ cumulative distribution functions $F_0, F_1,\ldots,F_s$ such that $F_j\neq F_{j+1}$, for $j\in\{0,\ldots,s-1\}$, and the cumulative density function of each $X_i$, $F_{X_i}$, can be written as:
$$F_{X_i} = \left\{\begin{array}{ll}
F_0&\text{if }1\leq i < \tau_{1,n}  \\ 
F_1&\text{if }\tau_{1,n}\leq i < \tau_{2,n}  \\ 
\vdots&  \\ 
F_s&\text{if }\tau_{s,n}\leq i \leq n 
\end{array}\right. $$
In this context we say that $\tau_1,\ldots,\tau_s$ are the change-points in $\{X_1,\ldots,X_n\}$.

The previous general formulation admits three important restrictions: 1) if the $F_1,\ldots,F_s$ are assumed to belong to the same location family then the associated CPP is called the {\it mean change-point problem} (mCPP); 2) if $F_1,\ldots,F_s$ belong to the same scale family then we have the called {\it variance change-point problem} (vCPP) and; 3) if the cumulative distribution functions belong to the same location-scale family then we have the called {\it mean and variance change-point problem} (mvCPP). Here we focus on the mCPP.

Assuming that $s$ is the number of change-points in a sequence the statistical problem of testing the existence of at least one change is equivalent to the following hypothesis test: $H_0:s = 0$; $H_1:s > 0$. If we are unable to reject $H_0$ than nothing else has to be done except find estimates for the desired quantities; otherwise, finding good estimates for the change-points is a necessary step for any downstream analysis.

Solutions to the CPP are of great importance in many areas of research such as climatology, finance and genetics and, for that reason, a great variety of solution methods exist. The study of change-point problems started in \cite{Page1} where a approach based on cumulative sums (CUSUM), further developed in \cite{Page2}, was proposed. After this initial study the change-point problem received a large amount of attention resulting in numerous methods and approaches, such as: use of the maxima of statistics calculated along the sequence of observations \citep{CBS2,MOSUM_esti,wildbs,brodskybook,wildbs}, optimization methods \citep{segneigh,PELT} and hidden Markov models \citep{segneigh,PELT,oncosnp, penncnv,quantisnp,beroukhim2006inferring,ha2012integrative}. Those examples do not fully encompass the extent of change-point analysis research. \cite{cpreview} provides a extensive review on various subfields of change-point analysis.

There are many areas of applied research that benefits from change-point analysis ethnics. In statistical processes control one is interest in detecting, as fast as possible, the point after which a processes stopped behaving in the expected way \citep{montgomery2009statistical}. It is clear that change-point analysis technics can be applied in detecting this type of change 
\citep{basseville1993detection}. Other area where change-point analysis plays a role is econometrics where it is important to detect changes in the behavior of time ordered data \citep{bai2003computation, allen2013nonparametric}.

In genetics the study of how and what characteristics of a individual's genetic content might contribute to the occurrence and evolution of cancer has fundamental importance in the diagnosis and treatment of such diseases. Of particular interest in this context are the copy-number structure of a genome.

In medicine research, specially genetics, change-point analysis alse plays a crucial role in the understand of complex diseases (diseases that are consequence of a complex interations between multiple genetical and environmental factors). The invention of array technologies enabled the analysis of a large number of DNA locus (with arrayCGH ou arraySNP) or even the whole genome ( with whole genome sequencing). Usually the results of a array experiment are two numerical sequences one of those related to the DNA copy-number structure. Changes in this sequence level indicates increments or decrements in the analyzed DNA genomic content. Those changes have been shown to effect the occurrence and development of cancers.

The evolution of technological capabilities makes the segmentation of such sequences even more challenging since modern methods generate  data with a large numbers of observations, possibly more than 3billion observations (one for each base in the DNA chain). Searching for change-points on such large data sets can only be feasible with methods whose computational complexity are as close as possible to linear. Another issue is that as the technological evolution increases the detection of small segments becomes less practically relevant, \cite{density}, and might produce unreliable results since a bigger false positive rate is inevitable if the chosen method do not prevent the detection of small segments. 

We propose an adaption to a existing method, \citep{sara1}, to the scenario where the large amount of data hinders the application of other slower methods, where the adoption of a minimum length for the distance between segments is reasonable and the sample distribution deviations from normality hinders the application of methods relying on the normal distribution.

$\{X_1,\ldots,X_n\}$ is a sequence of independent random variables and we take $\mu_i = \mathbb{E}(X_i)$ and $\sigma^2 = \mathbb{V}(X_i)<\infty$. We also adopt the following notation: $[i:j] = \{i,i+1,\ldots,j\}$ if $i$ and $j$ are natural numbers such that $i<j$. If $I\subset\{1,\ldots,n\}$ then
$$\overX_{I} = \frac{1}{|I|}\sum_{i \in I}X_i,$$
where $|I|$ denotes the cardinality of $I$.

\section{The SaRa Segmentation Procedures}

\cite{sara1} proposed a method based on the intuition that the property of a particular position $i$, along the sequence $\{X_1,\ldots,X_n\}$, being a change-point can be assessed by consulting a vicinity of $i$. Based on this simple observation the authors use the cumulative sum of neighboring regions:
$$L_i^k = \left|\sum_{j = i - k}^{i - 1}X_{j} - \sum_{j = i}^{i + k - 1}X_{j}\right|,$$
where $i = k + 1,\ldots,n - k + 1$ to quantify the evidence of the existence of a change-point in position $i$. We define $V(i) = [i-k,i + k - 1]$ for notation easy.

The procedure for calling change-points is based on two steps:
\begin{itemize}
	\item Screening step; change-point candidates are selected by finding the values of $i$ such that $L_{i}^k \geq L_{j}^k$ for all $j\in V(i)$ and $L_i^k > \delta$ where $\delta$ is a predefined real constant. That is, the method select all the global maxima in the vicinities $V(i)$ which are greater than a threshold $\delta$. The value of $\delta$ is suggested to be related to the minimal value of a shift in the mean. We will denote by $\mathcal{J}$ the set of all candidate change-points found in this step.
	\item Ranking step; The next step is the ranking of all candidate change-points on $\mathcal{J}$ with respect to some information criteria. In \cite{sara1} the authors adopt the Bayesian information criteria (BIC) and the following greed heuristic to select a final set of change-points is proposed. Let $\mathcal{J}_{(i)}$ be the result of removing the $i$-th candidate change-point from $\mathcal{J}$. If no set $\mathcal{J}_{(i)}$ has a lower BIC than $\mathcal{J}$ then $\mathcal{J}$ is the final set of change-points. Otherwise let $i$ be the index such that $\mathcal{J}_{(i)}$ is the set with minimal BIC value, then $\mathcal{J}$ is updated to $\mathcal{J}_{(i)}$ and we restart the second step with this reduced set of change-points candidates.
\end{itemize}

There are some issues in utilizing the SaRa algorithm as proposed by \cite{sara1}. A meaningful value of $\delta$ might not be easily selected because impurities in the cell samples from which the LRR sequences were generated change the minimal difference between neighboring segments means. Also the choice of BIC as a criteria for removing incorrectely identified change-point candidates make the method model dependent and deviations from the normal distributions might result in unwanted results. Other problems with the BIC criteria were identified in \cite{sara2}. 

\cite{sara2} propose improvements to overcome these difficulties. The authors propose the use of quantile normalization in a prepossessing step to address departures from normality. The selection of $\delta$ is done via a Monte Carlo simulation procedure for finding a particular quantile of the maximum of $L_i^k$ distribution. The authors also noted that the ranking procedure did not remove false positives in a satisfactory way. They then proposed a new procedure for removing incorrectly detected change-points. This new approach uses the clustering information of all segments given by the initial change-points candidates $\mathcal{J}$ in, at most, three groups with the help of the modified bayesian criteria. When two neighboring segments are assigned to the same cluster the change-point between them is removed from $\mathcal{J}$.

The modifications formulated by \cite{sara2} resulted in a strategy that is more robust to deviations from normality than the SaRa method. One disadvantage, noted by the authors, is that the number of clusters chosen to group the segments may affect the segmentation when the true number of cluster differ from the chosen value. Another issue is that the use of quantile normalization may unduly influence the analysis of microarray experiments data, as noted by \cite{qiu2013impact}. Another difficulty is the execution time burden of performing Monte Carlo methods for large samples.

In the next section we propose a strategy to select the threshold $\delta$ when we do not have any information regarding the minimum mean change required to identify a change-point. We also propose a novel approach, robust to sample distribution and size, to reduce the number of incorrectly detected change-points. Our procedures are best suited for situations where the minimal length of a segment is not so small (at least 25 observations). The length requirement is a sugestion made by microarray manufacturers, like Affymetrics, to reduce false positives on downstream analysis. Despite the requirement, this is a common situation, given that recent microarray products contain millions of probes \citep{density}. 

\section{Screening and Merging Segmentation Procedure}
We present a method closely related to the SaRa method. Our modifications are: the measure used to assess whether or not a position is a change-points, the choice of threshold $\delta$ and the procedure for removing incorrectly select change-points candidates.
\subsection{Threshold Selection and Screening Procedure}
We modify $L_i^k$ by defining the following statistic:
$$M_{i}^k=\frac{1}{S_{seq}}\left(\frac{2}{k}\right)^{-1/2}\left|\overX_{[i-k:i-1]}- \overX_{[i:i+k-1]}\right|,$$ 
where 
$$S_{seq}^2 = \frac{1}{2(n-1)}\sum_{i=2}^n(X_i - X_{i-1})^2,$$
is a consistent estimator of $\sigma^2$, even in the presence of multiple change-points, given a few weak conditions on the change-points evolution as $n$ increases.

We carried the screening step of SaRa in the same way but replacing the statistic $L_i^k$ with $M_i^k$. It allows a simple, intuitive and fast way of selecting a threshold $\delta$. 

If we assume that the $X_i$ are normally distributed (or that $k$ is big enough) and that there are no change-points along the sequence, then each $M_i^k$ is approximately distributed as a folded normal distribution, having cumulative distribution function $F_{fold}$ given by
\begin{equation}
F_{fold}(x) = \int_{0}^{x}\sqrt{\frac{2}{\pi}}e^{-u^2/2}d,u 
\end{equation}

We propose the use of the upper tail quantiles of $F_{fold}$, i.e., for $\alpha\in (0,1)$ take $\delta(\alpha) = F_{fold}^{-1}(1-\alpha)$ as the threshold for the screening.

As pointed by \cite{sara1}, the screening procedure can be with more than one value of $k$ and we also utilize this approach in our method. 

\subsection{Merging Procedure}
After a series of screening procedures is applied to $\{X_1,\ldots,X_n\}$, we are left with a set of change-point candidates, $\mathcal{J} = \{\hat{\tau}_{1},\ldots,\hat{\tau}_{p}\}$, which might contain change-points that do not satisfy the condition on the minimum segment length and candidates that represent false-positive detections. To filter those cases from $\mathcal{J}$ we adopt the following procedure: we sequentially test if the mean of observations on the left of $\hat\tau_i$ differs from the mean of observations on the right of $\hat\tau_i$. That is, for each $i$ we compute the usual statistic for comparing the mean of two groups:
$$
T = \frac{1}{S_{seq}}\left(\frac{1}{\hat{\tau}_i - \hat{\tau}_{i-1} + 1} + \frac{1}{\hat{\tau}_{i+1} - \hat{\tau}_{i}}\right)^{-1/2}(\hat{\mu}_{[\hat{\tau}_{i-1}:\hat{\tau}_{i}-1]} - \hat{\mu}_{[\hat{\tau}_{i}:\hat{\tau}_{i+1}-1]})
$$
and compare it to the critical value found by the normal approximation to the distribution of $T$.

If $T$ fall outside a predefined confidence interval then we proceed to the next change-point candidate $\hat{\tau}_{i+1}$. Otherwise we remove $\hat{\tau}_i$ from the list of change-points candidates and since this change the observations to the right of $\hat\tau_{i-1}$ we reposition $\hat\tau_{i-1}$ with
$$
\hat{\tau}_i = \underset{\tau_{i-1}+k'\leq j\leq \tau_{i+1}-k'+1}{\text{arg max}}\left(\frac{1}{j-\hat{\tau}_{i - 2}}+\frac{1}{\hat\tau_{i} - j}\right)^{-1/2}\left|\hat\mu_{[\hat\tau_{i-2}:\hat\tau_{i-1}-1]} - \hat\mu_{[\hat\tau_{i-1}:\hat\tau_{i}-1]}\right|,
$$
which is the estimator of a single change-point based on the likelihood ratio statistic for the case of normally distributed data \citep{sen_BS3}. Note that the value of $k'$ does not need equal $k$.

Because this methods is based on merging adjacent segments that fail to reject the hypothesis that they share the same mean we call this step the merge step. And, for the same reason, we call our method screening and merging (SaMe) segmentation method.

\section{Simulation Studies}\label{sim_sec}
We now turn to the issue of assessing the ability of our proposed methods to correctly identify the location of change-points in the mean along a sequence of observations. Also we are interested in how the stochastic distribution of said observations might affect the method's performance.

We compare our method SaMe to the well established procedures proposed in \cite{CBS2}, which we denote by CBS, and \cite{PELT}, which we denote by PELT.

The parameter chosen for our method are $\alpha = 0.01$ both in the screening step and in the merging step. We utilize a multiple bandwidths approach with $k \in \{25, 50, 100\}$ in the screening step. The value of $k'$ is the merge step is set to 20.

In the application of CBS we followed the authors recommendations adopting a level of significance of $0.01$. The parameters chosen for PELT will depend on the particular data distribution. In all situations we adopt a minimum distances between change-points of 20 (for consistency with the value of $k'$). For normally distributed data we chose the BIC penalization (since choosing AIC would result in too large false-positives detection) and in the case where we sample observations from a real data set we chose the AIC penalization (since in this case the BIC alternative result in a very low rate of change-point detection).

The criteria for accessing how fast a method estimate the change-points was its execution time in seconds (t). The accuracy of a method is accessed in the following way, if $\tau$ is the index associated with a change-point we consider two windows containing $\tau$: $W^{tol}_\tau = \{\tau-tol,\ldots,\tau,\ldots,\tau+tol\}$ where $tol$ is ether $10$ or $5$. We say that a method detected $\tau$ with tolerance $tol$ if there is a estimated change-point inside $W^{tol}_\tau$. We denote by $p10$ the percentage of change-points that where detected with tolerance $10$ by the method. Also $p5$ is the analogous with tolerance $5$. The last criteria is FP denoting the mean difference between the actual number of change-points and the number of real change-points detected with tolerance 10. This last criteria is a way of accessing the rate of false discoveries of change-points.  
\subsection{Normaly Distributed Data}
In the first set of simulations we consider a sequence of observations $\{X_1,\ldots,X_n\}$ with $n = 10,000$ and such that
$$X_i = f(i) + \epsilon_i,$$
where the $\epsilon_i$ are all $N(0,1)$ and the function $f$ specify the change-point structure. To fully investigate the impact of change-points positions on the results of the various methods tested the form of $f$ for each particular simulation is chosen in the following way. First the number of change-points $m$ along the sequence is fixed to be one of $2$, $4$ or $6$, the length of the segment determined by the change-points $l$ is chosen to be on of the values $25$, $50$ or $100$, finally the mean shift $s$ is fixed at one the values $1$, $1.5$ or $2$. Then we randomly sample $m/2$ natural numbers, $cp_1,\ldots,cp_{m/2},$ uniformly form $[(i-i)L + l:i*L - l)$, where $L$ is the greatest integer smaller than $10000/(m/2)$. The change-points positions are chosen as $\{cp_1, cp_1 + l, \ldots, cp_{m/2}, cp_{m/2} + l\}$ and $f(i)$ is given by $f(x) = s$, if $cp_i\leq i < cp_i + l$, for some $i\in\{1,\ldots,m/2\}$, and $f(x) = 0$ elsewhere. For all the $27$ possible combinations of $m$, $l$ and $s$ we construct $100$ random samples.

The first thing to note is that all methods perform well in terms of p5 and p10 when $s = 1.5$ or $s = 2$, and for this reasons those results are omitted. It is not surprising since the presence of a change-point becomes more evident the greater $s$ is. For the values of $s = 1$ there are differences between the methods.

\setlength\extrarowheight{1.5pt}
\begin{table}[htb]
	\begin{center}
		\caption{\label{shift1}Comparison of SaMe, CBS and PELT on normally distributed data with changes in mean of magnitude 1. In terms of execution time and change-point detection SaMe outperformed both CBS and PELT. SaMe and CBS have similar false-discovery rates in this case. PELT have the smallest false-discovery rate.}
		\begin{tabular}{ c c    C{2cm}   C{2cm}   C{2cm} }
			&  & SaMe & CBS & PELT \\\hline
			\multirow{4}{*}{ single } & t & 0.1532 & 0.8936 & 0.4226 \\
			& p10 & 0.5400 & 0.4500 & 0.2700 \\
			& p5 & 0.4900 & 0.3700 & 0.2400 \\
			& FP & 0.0700 & 0.0600 & 0.0200 \\ \hline
			\multirow{4}{*}{ double } & t & 0.1386 & 1.0853 & 0.3106 \\
			& p10 & 0.5500 & 0.5125 & 0.2675 \\
			& p5 & 0.5050 & 0.4625 & 0.2575 \\
			& FP & 0.0900 & 0.1100 & 0.0500 \\ \hline
			\multirow{4}{*}{ trip } & t & 0.1317 & 1.4698 & 0.2352 \\
			& p10 & 0.5100 & 0.5183 & 0.2317 \\
			& p5 & 0.4717 & 0.4633 & 0.2167 \\
			& FP & 0.1100 & 0.2900 & 0.0500 \\ \hline
		\end{tabular}
\end{center}
\end{table}
			
\begin{table}[htb]
	\begin{center}
		\caption{\label{shift2}Comparison of SaMe, CBS and PELT on normally distributed data with changes in mean of magnitude 1. SaMe has the smallest execution time. In terms of change-point detection CBS outperforms SaMe and PELT in all instances. False discoveries rates of all methods are similar.}
		\begin{tabular}{ c c    C{2cm}   C{2cm}   C{2cm} }
			&  & SaMe & CBS & PELT \\\hline	
			\multirow{4}{*}{ single } & t & 0.1400 & 0.6906 & 0.3823 \\
			& p10 & 0.8100 & 0.9200 & 0.8200 \\
			& p5 & 0.7250 & 0.7950 & 0.7050 \\
			& FP & 0.2500 & 0.1600 & 0.1200 \\ \hline
			\multirow{4}{*}{ double } & t & 0.1330 & 0.8445 & 0.2669 \\
			& p10 & 0.8900 & 0.9150 & 0.8500 \\
			& p5 & 0.7975 & 0.8050 & 0.7450 \\
			& FP & 0.2000 & 0.4100 & 0.2600 \\ \hline
			\multirow{4}{*}{ trip } & t & 0.1105 & 0.7965 & 0.2042 \\
			& p10 & 0.8717 & 0.9367 & 0.8717 \\
			& p5 & 0.7850 & 0.8417 & 0.7600 \\
			& FP & 0.3300 & 0.4200 & 0.3200 \\ \hline
		\end{tabular}
\end{center}
\end{table}			
			
\begin{table}[htb]
	\begin{center}
		\caption{\label{shift3}Comparison of SaMe, CBS and PELT on normally distributed data with changes in mean of magnitude 1. All methods have similar change-point detection ability and false-positive rates. SaMe once again has the smallest execution time.}
		\begin{tabular}{ c c    C{2cm}   C{2cm}   C{2cm} }	
			&&\multicolumn{3}{c}{size = 100}\\
			\multirow{4}{*}{ single } & t & 0.1372 & 0.5050 & 0.3819 \\
			& p10 & 0.8750 & 0.9250 & 0.9500 \\
			& p5 & 0.7450 & 0.8450 & 0.8300 \\
			& FP & 0.3800 & 0.1900 & 0.1000 \\ \hline
			\multirow{4}{*}{ double } & t & 0.1195 & 0.5420 & 0.2668 \\
			& p10 & 0.9300 & 0.9350 & 0.9400 \\
			& p5 & 0.7875 & 0.8225 & 0.8450 \\
			& FP & 0.3300 & 0.3500 & 0.2400 \\ \hline
			\multirow{4}{*}{ trip } & t & 0.1019 & 0.5903 & 0.1918 \\
			& p10 & 0.9383 & 0.9600 & 0.9367 \\
			& p5 & 0.8200 & 0.8500 & 0.8383 \\
			& FP & 0.4500 & 0.4400 & 0.3900 \\ \hline
		\end{tabular}
	\end{center}
\end{table}

Table \ref{shift1} shows that for the case of a single altered segments of size 25 and mean shift of magnitude 1 the SaMe method outperform all other methods in terms of the number of correctely identified change-points. When there are two or three altered segments SaMe still is the best methods in terms of change-point identification ability, but CBS performs very close to it. PELT had a poor change-point identification performance for every number of altered segments. In terms of execution time CBS took the longest to complete execution. SaMe was by far the fastest method. False discovery rates were similar for CBS and SaMe while PELT produced the lest number of false-discoveries.

In Table \ref{shift2} we see the results for the case of altered segments of length 50 and mean shift difference of magnitude 1. In this case we can see a slight increase in false-discovery rates for CBS and SaMe. The execution time follow the same patter with SaMe been the fastest. In terms of detection ability CBS was the most accurate method. SaMe and PELT had a similar detection ability.

Table \ref{shift3} show the results for the case of altered segments of length 100 and mean shift difference of magnitude 1. The results here are similar to the ones obtained in Table \ref{shift2} but now the detection abilities of each methods are closer to one another.

\subsection{Sampling Real Data}
We perform a second simulations study since the critical values for SaMe are approximations based on the normal distribution. This allow us to evaluate the effects of deviation from normality. We perform a simulation study using observations sampled from a pool of data arising from a experiment in genetics. The data is available in the \texttt{acnr} R package \citep{acnr}.

In a similar way as was done in the previous simulation experiment we generate sequences $\{X_1,\ldots,X_n\}$ but in this case if $\{cp_1, cp_1 + l,\ldots, cp_{m/2}, cp_{m/2} + l\}$ are the randomly selected change-points, then if $cp_j \leq i < cp_j + l$ for some $j\in\{1,\ldots,m/2\}$ then $X_i$ is drawn from a region with a deletion or a simple-gain in one allele (this is a exclusive "or"), otherwise is drawn from a copy-number neutral sample. The data set chosen for our simulations is labeled "GSE29172" in \texttt{acnr}.

Tables \ref{real1}, \ref{real2}, \ref{real3}, \ref{real4}, \ref{real5}, \ref{real6}. Some of the general observations made in the last subsection remains true here. In all stances SaMe is the fasts method by a large margin and in terms of false-discovery rate SaMe and CBS have a similar performance. PELT have a very small number of false-discovery but this is mainly due to its inability to detect small segments. 

When the altered segments were sample from the copy-number 1 dataset all methods are able to detect more than 95\% of the present change-points. But when the altered change-points are from a dataset with copy-number 3 Table \ref{real4} shows that CBS has a slight better ability to detect change-points in relation to SaMe, while PELT detect only a small percentage of those changes. In Tables \ref{real5} and \ref{real6} we see that when the altered segments are of size 50 or greater all methods detect more than 95\% of the present change-points.
\begin{table}[htb]
	\begin{center}
		\caption{\label{real1}Comparison of SaMe, CBS and PELT when altered segments are of copy number 1 and of length 25. SaMe has the fastest execution time while CBS has a slight better change-point detection ability and PELT produce the smallest number of false-discovery.}
		\begin{tabular}{ c c    C{2cm}   C{2cm}   C{2cm} }
			&  & SaMe & CBS & PELT \\\hline
			\multirow{4}{*}{ single } & t & 0.0867 & 0.5887 & 0.4353 \\
			& p10 & 0.9500 & 0.9900 & 0.9500 \\
			& p5 & 0.9300 & 0.9750 & 0.9450 \\
			& FP & 0.2200 & 0.2300 & 0.0000 \\ \hline
			\multirow{4}{*}{ double } & t & 0.0769 & 0.6212 & 0.2840 \\
			& p10 & 0.9875 & 1.0000 & 0.9275 \\
			& p5 & 0.9775 & 0.9950 & 0.9200 \\
			& FP & 0.1300 & 0.1600 & 0.0000 \\ \hline
			\multirow{4}{*}{ trip } & t & 0.0709 & 0.6473 & 0.2243 \\
			& p10 & 0.9700 & 1.0000 & 0.9267 \\
			& p5 & 0.9517 & 0.9933 & 0.9200 \\
			& FP & 0.1800 & 0.1700 & 0.0000 \\ \hline
		\end{tabular}
	\end{center}
\end{table}

\begin{table}[htb]
	\begin{center}
		\caption{\label{real2}Comparison of SaMe, CBS and PELT when altered segments are of copy number 1 and of length 50. SaMe has the fastest execution, PELT produce the smallest number of false-discovery and all methods detect all present change-points.}
		\begin{tabular}{ c c    C{2cm}   C{2cm}   C{2cm} }
			&  & SaMe & CBS & PELT \\\hline
			\multirow{4}{*}{ single } & t & 0.0893 & 0.6596 & 0.4597 \\
			& p10 & 1.0000 & 1.0000 & 1.0000 \\
			& p5 & 0.9950 & 0.9950 & 0.9950 \\
			& FP & 0.2500 & 0.1600 & 0.0000 \\ \hline
			\multirow{4}{*}{ double } & t & 0.0750 & 0.6012 & 0.2763 \\
			& p10 & 1.0000 & 1.0000 & 1.0000 \\
			& p5 & 0.9925 & 0.9925 & 0.9950 \\
			& FP & 0.2200 & 0.1900 & 0.0000 \\ \hline
			\multirow{4}{*}{ trip } & t & 0.0733 & 0.6312 & 0.2201 \\
			& p10 & 1.0000 & 1.0000 & 1.0000 \\
			& p5 & 0.9917 & 0.9950 & 0.9933 \\
			& FP & 0.2200 & 0.2700 & 0.0000 \\ \hline
		\end{tabular}
	\end{center}
\end{table}

\begin{table}[htb]
	\begin{center}
		\caption{\label{real3}Comparison of SaMe, CBS and PELT when altered segments are of copy number 1 and of length 100. Similar results to the previous Table \ref{real2}.}
		\begin{tabular}{ c c    C{2cm}   C{2cm}   C{2cm} }
			&  & SaMe & CBS & PELT \\\hline
			\multirow{4}{*}{ single } & t & 0.0870 & 0.5656 & 0.4160 \\
			& p10 & 1.0000 & 1.0000 & 1.0000 \\
			& p5 & 1.0000 & 0.9900 & 0.9850 \\
			& FP & 0.4000 & 0.1000 & 0.0000 \\ \hline
			\multirow{4}{*}{ double } & t & 0.0742 & 0.5435 & 0.2712 \\
			& p10 & 1.0000 & 1.0000 & 1.0000 \\
			& p5 & 0.9900 & 0.9875 & 0.9925 \\
			& FP & 0.2400 & 0.1800 & 0.0000 \\ \hline
			\multirow{4}{*}{ trip } & t & 0.0752 & 0.6267 & 0.2169 \\
			& p10 & 0.9967 & 1.0000 & 1.0000 \\
			& p5 & 0.9933 & 0.9917 & 0.9883 \\
			& FP & 0.2400 & 0.1700 & 0.0000 \\ \hline
		\end{tabular}
	\end{center}
\end{table}

\begin{table}[htb]
	\begin{center}
		\caption{\label{real4}Comparison of SaMe, CBS and PELT when altered segments are of copy number 3 and of length 25. SaMe has the fastest execution time while CBS has a better change-point detection ability spatially for the case where only one altered segment is present. PELT produce the smallest number of false-discovery and was particularly unable to detect change-points.}
		\begin{tabular}{ c c    C{2cm}   C{2cm}   C{2cm} }
			&  & SaMe & CBS & PELT \\\hline
			\multirow{4}{*}{ single } & t & 0.0854 & 0.7368 & 0.4577 \\
			& p10 & 0.7700 & 0.8450 & 0.3050 \\
			& p5 & 0.7400 & 0.7700 & 0.3000 \\
			& FP & 0.3100 & 0.2100 & 0.0100 \\ \hline
			\multirow{4}{*}{ double } & t & 0.0852 & 1.0256 & 0.3019 \\
			& p10 & 0.8800 & 0.8725 & 0.3450 \\
			& p5 & 0.8550 & 0.8200 & 0.3325 \\
			& FP & 0.1400 & 0.1900 & 0.0400 \\ \hline
			\multirow{4}{*}{ trip } & t & 0.0802 & 1.0088 & 0.2400 \\
			& p10 & 0.8350 & 0.8850 & 0.3617 \\
			& p5 & 0.7867 & 0.8183 & 0.3517 \\
			& FP & 0.3300 & 0.3400 & 0.0300 \\ \hline
		\end{tabular}
	\end{center}
\end{table}

\begin{table}[htb]
	\begin{center}
		\caption{\label{real5}Comparison of SaMe, CBS and PELT when altered segments are of copy number 3 and of length 50. SaMe has the fastest execution time and CBS has a slightly better change-point detection ability. PELT produce the smallest number of false-discovery.}
		\begin{tabular}{ c l   | C{2cm}  | C{2cm}  | C{2cm} }
			&  & SaMe & CBS & PELT \\\hline
			\multirow{4}{*}{ single } & t & 0.0862 & 0.6367 & 0.4112 \\
			& p10 & 0.9550 & 0.9800 & 0.9550 \\
			& p5 & 0.8800 & 0.8950 & 0.8550 \\
			& FP & 0.4700 & 0.2200 & 0.0300 \\ \hline
			\multirow{4}{*}{ double } & t & 0.0837 & 0.5872 & 0.2651 \\
			& p10 & 0.9725 & 0.9950 & 0.9475 \\
			& p5 & 0.9175 & 0.9200 & 0.9050 \\
			& FP & 0.3400 & 0.2300 & 0.1300 \\ \hline
			\multirow{4}{*}{ trip } & t & 0.0731 & 0.6653 & 0.2263 \\
			& p10 & 0.9733 & 0.9783 & 0.9717 \\
			& p5 & 0.9117 & 0.9200 & 0.9183 \\
			& FP & 0.3300 & 0.4000 & 0.0700 \\ \hline
		\end{tabular}
	\end{center}
\end{table}

\begin{table}[htb]
	\begin{center}
		\caption{\label{real6}Comparison of SaMe, CBS and PELT when altered segments are of copy number 3 and of length 100. SaMe has the fastest execution time. With respect of other criteria the methods have a similar behavior.}
		\begin{tabular}{ c l   | C{2cm}  | C{2cm}  | C{2cm} }
			&  & SaMe & CBS & PELT \\\hline
			\multirow{4}{*}{ single } & t & 0.0870 & 0.6285 & 0.3817 \\
			& p10 & 0.9800 & 0.9900 & 0.9750 \\
			& p5 & 0.9050 & 0.9300 & 0.9000 \\
			& FP & 0.2300 & 0.1300 & 0.0500 \\ \hline
			\multirow{4}{*}{ double } & t & 0.0797 & 0.6002 & 0.2822 \\
			& p10 & 0.9775 & 0.9850 & 0.9875 \\
			& p5 & 0.9275 & 0.9275 & 0.9225 \\
			& FP & 0.2600 & 0.2100 & 0.0500 \\ \hline
			\multirow{4}{*}{ trip } & t & 0.0728 & 0.6383 & 0.2265 \\
			& p10 & 0.9817 & 0.9833 & 0.9883 \\
			& p5 & 0.9217 & 0.9217 & 0.9350 \\
			& FP & 0.3100 & 0.3300 & 0.0700 \\ \hline
		\end{tabular}
	\end{center}
\end{table}

\section{Real Data Examples}
In this section we show how the SaMe and other methods perform on real datasets. The first dataset we consider, available at \texttt{http://penncnv.openbioinformatics.org}, is comprised of a father-mother-offspring trio produced with Illumina 550K platform. Following the analysis in \cite{sara1} we focus on the offspring log-R-ratio sequence for chromosomes 3, 11 and 20. To each of these sequences we apply SaMe, CBS and PELT. The parameters for SaMe are $k = 25$, $k' = 20$ and pValues of $0.01$ for any test performed. We choose CBS parameters in the same way we did in the previous section with the exception that here we adopt a change-point pruning procedure that removes a change-point. PELT parameters are the same ones as in the previous section and we consider separately the BIC and AIC criterions.

\setlength\extrarowheight{1.5pt}
\begin{table}[htb]
	\begin{center}
		\caption{\label{penncnv_cp} Comparison of the number of change-points detected by SaMe, CBS and PELT for three chromosomes of the offspring dataset.}
		\begin{tabular}{ C{3cm} |  C{2cm}   C{2cm}   C{2cm} C{2cm} }
			& SaMe  & CBS &  PELT(BIC) & PELT(AIC) \\\hline
		 Chromosome 3 &  2    & 24  & 0 & 2\\  
			Chromosome 11 & 5 &  12   & 2 &2\\
			Chromosome20 & 2   & 12   & 0 & 0
		\end{tabular}
	\end{center}
\end{table}

Table \ref{penncnv_cp} shows the number of change-points detected by each method. Our results are very similar to those found in \cite{sara1} where the execution of \texttt{PennCNV} \cite{penncnv} found 2, 4 and 2 change points along chromosomes 3, 11 and 20, respectively. Because \texttt{PennCNV} is a well established change-point detection software it is safe to assume that methods whose performance are close that of \texttt{PennCNV} produced meaningful results. In this experiment SaMe has the closest result to that of \texttt{PennCNV}, giving some indication that it can perform well on genetic data.

\setlength\extrarowheight{1.5pt}
\begin{table}[htb]
	\begin{center}
		\caption{\label{penncnv_time}Comparison of the execution time (in seconds) for SaMe, CBS and PELT on three chromosomes of the offspring dataset.}
		\begin{tabular}{ C{3cm} |  C{2cm}   C{2cm}   C{2cm} C{2cm} }
			& SaMe  & CBS &  PELT(BIC) & PELT(AIC) \\\hline
			Chromosome 3 &  1.6622    & 7.3892  & 8.6240 & 8.2734\\  
			Chromosome 11 & 0.3760 &  2.6015   & 2.3227 &2.2863\\
			Chromosome20 & 0.2843   & 2.2871   & 1.2374 & 1.2034
		\end{tabular}
	\end{center}
\end{table}

Table \ref{penncnv_time} presents the execution time for each method on each chromosome. The results here partially reproduce those found in the previous section. Here SaMe constantly outperformed the other methods by a large margin. We can see that for chromosome 3 CBS was faster than PELT. 

Our second real data study is based on a dataset composed of 482 samples of normal and tumoral cells, from which 29 samples are available in \texttt{https://www.ncbi.nlm.nih.gov/geo}, of 259 men in a study regarding prostate cancer \citep{sample}.  

We apply SaMe, CBS, PELT and \texttt{oncoSNP} to the 24 tumoral available samples. \texttt{oncoSNP} is one of the newest segmentation software and is based on a hidden Markov model that consider simultaneously both sequences (log-R-ratio and B-allele-frequency) resulting from a SNP-array experiment. It is available at \texttt{https://sites.google.com/site/oncosnp/}. We set the parameters of SaMe at $\alpha = 0.01$ in both the screening and the merging procedures, $k \in \{25,50,100\}$ and $k' = 20$. We chose CBS parameters in the same way we did in the previous analysis. PELT parameter's also are fixed in the same way as the last study with the BIC penalty. \texttt{oncoSNP} is parametrized following the authors guidelines.

Following the analysis proposed in \cite{sara2} we observed that approximately 95\% of the segments identified, by \texttt{oncoSNP}, with copy-number alterations are smaller than 2163. Also 90\% of all segments are smaller than 2163. Those values are good guidelines to evaluate a method performance. Resulting segmentations with percentages close to the ones presented indicate the existence of meaningful segments.

After applying SaMe, CBS and PELT to all 24 tumoral samples we find that among the segmentations resulted in approximately 97\%, 99\% and 41\% of the segments smaller than 2163, respectively. Also the number of segments identified by \texttt{oncoSNP}, SaMe, CBS and PELT is 31,403, 137,654, 285,842,  6,755, respectively. Those results indicate that although SaMe and might produce some false-positives it does so to a much lesser extent than does CBS. Also PELT seens to produce a unreliable segmentation given the number of segments found and their length being too big. 

Since the results of oncoSNP can not be considered as truth this comparison is admittedly not ideal. However it gives some indication of how SaMe behaves in comparison to other methods.

\section{Discussion} 
In this article, we proposed a new version of the SaRa method where the threshold values are selected via a normal approximation and the control for erroneous change-point detections is made by a sequence of hypothesis tests on the first set of change-point candidates. This procedures avoids the necessity of selecting a information criteria and renders the method more robust to deviations from the normal distribution.

The results simulation studies we performed show that SaMe has a change-point detection ability similar to that of CBS and PELT, two well established methods of change-point detection, both on normally distributed observations or when observations are sampled from real datasets. Those experiments also show the robustness of our segmentation to the underling distribution. The same experiments also show that SaMe performs at a much higher speed in comparison to CBS and PELT. 

In real data studies we can see the indication that SaMe presents results in line with \texttt{PennCNV} and \texttt{oncoSNP} two two well established change-point detection method specially designed to SNP-array data.

The parameters necessary for the methods execution are: $\{k_1,\ldots,k_K\}$, the window sizes used in the screening step; $\alpha$, the p-value used in the construction of thresholds for each $k_i$; $k'$, the minimum size of a segment; and $\alpha'$ the p-value associated with the merging step. The values of $k_1,\ldots,k_K$ and $k'$ are not difficult to choose and are connected to the minimum number of observations allowed in a segment. The values of $\alpha$ and $\alpha'$ can be select from the standard $0.05$, $0.025$ and $0.01$ usually adopt for hypothesis testing.

Because our method performs a series of hypothesis tests along the sequence of observations the values of $\alpha$ and $\alpha'$ are not the actual significance level of the overall procedure. This issue seams not the be a problem since our simulation studies show that the number of generated false-positives is not alarming and is comparable to that of CBS.

Because there are a variety of SNP-array platforms, resulting in a set of possible data distributions, and because the technological evolution is producing datasets with large number of observations SaMe is a method to be considered in the analysis of this kind of data given its fast execution time and its tolerance to the underlying sample distribution.

\section{Acknowledgment}
We want to thank FAPESP (grant 2013/00506-1) and CAPES for providing the resources for the realization of this project.

\bibliographystyle{apalike}
\bibliography{biblio_LRR_2.bib} 

\end{document}